\documentclass[aps,amssymb,eqsecnum,showkeys,showpacs]{revtex4}

\usepackage{graphicx}
\usepackage{psfig,epsfig}
\newcommand{\be}{\begin{eqnarray}}
\newcommand{\ee}{\end{eqnarray}}
\newcommand{\bx}{{\bf x}}

\newcommand{\bz}{{\bf z}}
\newcommand{\br}{{\hat {\bf r}}}
\newcommand{\bs}{{\hat {\bf s}}}
\begin{document}


\title{Fermions on the light front transverse lattice}
\author{Dipankar Chakrabarti}
\email{dipankar@theory.saha.ernet.in}
\author{Asit K. De}
\email{de@theory.saha.ernet.in}
\author{A. Harindranath}
\email{hari@theory.saha.ernet.in}
\affiliation{
Theory Group, Saha Institute of Nuclear Physics,
1/AF, Bidhannagar, Kolkata 700064 India}

\date{17 November 2002}

\begin{abstract}
We  address the problems of fermions in light front QCD on a
transverse lattice.  We propose and numerically investigate different 
approaches of formulating fermions on the light front transverse lattice.
In one approach we use  forward and  backward derivatives. There is no 
fermion doubling and the helicity flip  term proportional to the fermion 
mass in the full light front QCD becomes an irrelevant term in the free 
field limit.  In the second approach with symmetric derivative (which has 
been employed previously in the literature), doublers appear and their 
occurrence is due to the decoupling of even and odd lattice sites. 
We study their removal from the  spectrum in two ways namely, light front 
staggered formulation and the  Wilson fermion formulation. The numerical
calculations in free field limit are carried out with both fixed  and
periodic boundary conditions on the transverse lattice and finite
volume effects are studied. We find that  an even-odd helicity flip
symmetry on the light front transverse lattice is relevant for fermion
doubling.
\end{abstract}

\maketitle
\section{Introduction}
Light front Hamiltonian formulation of transverse lattice QCD
\cite{Bardeen:1976tm,Bardeen:1980xx} has many interesting
features. With the gauge choice $A^{+}=A^0+A^3=0$ and the elimination of the
constrained variable $A^-=A^0-A^3$, it uses minimal gauge degrees of freedom in a 
manifestly gauge invariant formulation exploiting the residual gauge
symmetry in this gauge. So far encouraging results have been obtained
in the pure gauge sector and in the meson sector with 
particle number truncation (for a recent review see, Ref. \cite{review}).

It is well known that fermions on the lattice pose challenging
problems due to the doubling phenomenon. Light front formulation of
field  theory has its own peculiarities concerning fermions because of
the presence of a constraint equation. As an example, the usual chiral
transformation on the four component fermion field is incompatible
with the constraint equation for nonzero fermion
mass \cite{{Wilson:1994fk}}.  There have been previous studies of fermions on the
transverse lattice \cite{stagger,BK,dalme, buseal}. Our approach in this work is quite 
extensive and aims to understand  the  origin of the doublers.  We
identify an even-odd helicity flip symmetry of the light front
transverse lattice Hamiltonian, absence of which
means removal of doublers in all the cases we have studied. This is
closely related to the need to break chiral symmetry explicitly in the 
usual Euclidean formulation of lattice fermions.
  
As we shall see later in this article, 
the presence of the constraint equation in light front
field theory allows different methods to put fermions on a transverse
lattice. It is worthwhile to study all the different methods in order
to examine their strengths and weaknesses. 
In this work, we carry out a detailed numerical investigation of
three methods in free field limit with special emphasis on finite volume 
effects. We also
study the effects of imposing fixed and periodic boundary conditions
which have significant effects in finite volumes. There are two
important reasons to thoroughly study finite volume effects. First,
for a reasonable size of Fock space, computing limitations will force
us to be in a reasonably small volume when we deal with realistic
problems. Second, the currently practiced version of the transverse
lattice gauge theory uses {\it linear} link variables and 
recovering continuum physics is nontrivial. Finite volume studies are
also important in this connection.

In one of the approaches of treating fermions on the light front transverse
lattice,  we maintain as much transverse locality as possible on 
the lattice by using forward and backward derivatives without spoiling the 
hermiticity of the Hamiltonian. In this case
doublers are not present and the helicity flip term proportional
to the fermion mass in the full light front QCD becomes an irrelevant term  in the
free field limit. Thus in finite volume, 
depending on the boundary condition used, the two helicity states 
of the fermion may not be degenerate in the free field limit. However, 
we find that in the infinite volume limit the degeneracy is restored 
irrespective of the boundary condition. 

In the second approach \cite{BK}, symmetric derivatives are used 
which results  in a Hamiltonian  with only
next to nearest neighbor interaction when we take the free field limit. 
As a consequence even and odd
lattice sites decouple and the fermions live independently of each other on the
two  sets of sites. As a result we get four species of
fermions  on a two dimensional lattice  as excitations around 
zero transverse momentum (Note that this is quite
different from what one gets in the conventional Euclidean lattice theory when
one uses symmetric derivatives. In that case, doublers have at least one
momentum component near the edge of the Brillouin zone.). 
The doublers can be removed in more than one way.  We propose to use the 
staggered fermion formulation on the light front transverse lattice to eliminate 
two doublers and reinterpret the remaining two as  two
flavors.  In this  light front staggered fermion formulation, there is 
no flavor mixing in free field limit. But, in QCD, we get  irrelevant
flavor mixing terms.
An alternative which removes  doubling completely is to add the conventional Wilson 
term which generates many irrelevant interactions on the transverse
lattice. Among them, the  helicity flip interactions vanish but
the helicity non flip interactions survive in the free field limit.  
 

The plan of this paper is as follows. Notation  and  conventions are
presented in Sec. II. QCD Hamiltonian with forward-backward derivative is discussed
 and free  field limit is studied 
 in Sec. III.  QCD Hamiltonian with symmetric derivative with its
 free  field limit is considered 
 in Sec. IV.  
Staggered formulation and reinterpretation of doublers are discussed in Sec. V. 
 Removal of doublers  via the Wilson term is
studied in Sec. VI.   We discuss the even-odd spin flip symmetry and
its relation to
the fermion doubling on the light front transverse lattice in Sec. VII.
Finally Sec. VIII
contains summary and conclusions.
 In appendix A we compare and contrast the forward-backward derivative 
in conventional lattice  and light front transverse lattice for free
fermion field theory.
\section{Light Front Preliminaries}
\subsection{Notation and conventions}

The light front coordinates are $x^\pm = x^0 \pm x^3$, $x^\perp =
(x^1,x^2)$, the partial derivative $\partial^\pm = 2 {\partial \over
\partial x^\mp}$, the gamma matrices $\gamma^\pm = \gamma^0 \pm
\gamma^3$ and  projection operators 
$\Lambda^\pm = { 1 \over 4}\gamma^\mp \gamma^\pm$ . $x^+$ is the light 
front time and $x^-$ is the light front longitudinal coordinate.

The Lagrangian density for the free fermion is
\be
{\cal L}_{free} = {\bar \psi}( i \gamma^\mu \partial_\mu -m ) \psi.
\ee
Going to light front coordinates and 
using $ \psi^{\pm} = \Lambda^\pm \psi$,
\be
{\cal L}_{free} = {\psi^+}^\dagger i \partial^- \psi^+
+{\psi^-}^\dagger i \partial^+ \psi^- - {\psi^-}^\dagger (i
\alpha^\perp \cdot \partial^\perp + \gamma^0 m) \psi^+ -
{\psi^+}^\dagger( i \alpha^\perp \cdot \partial^\perp + \gamma^0 m) \psi^- ~. 
\ee
One of the  equations of motion from the above free Lagrangian is 
\be
i \partial^+ \psi^- = (i \alpha^\perp \cdot \partial^\perp + \gamma^0
m) \psi^+ ,\label{ce}
\ee
which is  actually a constraint equation because of absence of a time
derivative.  

$\psi^+$ is the dynamical fermion field and its equation of motion is
given by
\be
i \partial^- \psi^+ = (i \alpha^\perp \cdot \partial^\perp + \gamma^0
m) \psi^- .\label{de}
\ee
One can remove $\psi^-$ from Eq. (\ref{de}) using the constraint
equation (Eq. (\ref{ce})).  
The dynamical field $\psi^+$ can essentially be represented by two
components \cite{hz} such  that 
\be
\psi^+(x^-, x^{\perp}) = \left[ \begin{array}{l} \eta(x^-, x^{\perp})
\\ 0 \end{array} \right] ,
\ee
where $\eta$ is a two component field.  Its Fock expansion in the
light front quantization with tranverse directions discretized on a
two dimensional square lattice  is given by 
\be
\eta_{}(x^-, \bx) = \sum_\lambda \chi_\lambda \int {dk^+ \over 2 (2 \pi)
\sqrt{k^+}} \big [
b_{}(k^+,\bx,\lambda) e^{-{i \over 2} k^+ x^-} + d^\dagger_{}(k^+, \bx,
-\lambda) e^{{i \over 2}k^+x^-} \Big ],  
\ee
where $\chi_{\lambda}$ is the Pauli spinor, $\lambda=1,2$ denotes two
helicity states. $\bx$ now denotes the transverse
lattice  points. 

The canonical commutation relations are 
\be \{ b_{}(k^+,\bx, \lambda), b^\dagger_{}({k'}^+, \bx',\lambda') \} 
&=& \{ d_{}(k^+, \bx, \lambda),
d^\dagger_{}({k'}^+, \bx', \lambda')\} \nonumber \\
&=& 2 (2 \pi) k^+ ~\delta (k^+-{k'}^+) ~
\delta_{\lambda \lambda'} ~ \delta_{\bx, \bx'} ~ .
\ee
Using $
\int_{- \infty}^{+ \infty}dk^+ e^{{i \over 2} k^+(x^- - y^-)} = 2 (2 \pi)
\delta(x^- - y^-)$, we have,
\be
\Big \{ \eta_{}(x^-, \bx), \eta^\dagger_{}(y^-, \bx') \Big \} = 
\openone ~ \delta_{\bx, \bx'}~
 \delta(x^- - y^-)
\ee
where $\openone$ is a 2 by 2 unit matrix.

We use  Discretized Light Cone Quantization (DLCQ) \cite{BPP} for the
longitudinal dimension ($ -L \le x^- \le +L$) and implement
anti periodic boundary condition to avoid zero modes. Then,
\be
\eta(x^-, \bx) = { 1 \over \sqrt{2L}} \sum_\lambda \chi_\lambda
\sum_{l=1,3,5, \dots} [ b(l, \bx, \lambda)e^{-i { \pi l x^- / (2
L)}} + d^\dagger(l, \bx, -\lambda) e^{i { \pi l x^- / (2 L)}}] \label{fe}
\ee
with 
\be
\{ b (l, \bx, \lambda), b^\dagger(l', \bx', \lambda') \} =
 \{ d (l, \bx, \lambda), d^\dagger(l', \bx', \lambda') \} = 
\delta_{l l'} \delta_{\bx, \bx'} ~ \delta_{\lambda,
\lambda'}.
\ee
 In DLCQ with antiperiodic boundary condition, it is usual to
multiply the Hamiltonian $P^-$ by ${\pi \over L}$, 
so that $H={\pi \over L}P^-$ has the dimension of mass squared.

 In the following, for notational convenience we suppress $x^-$ in the
arguments of the fields.
\section{Hamiltonian with forward and backward derivatives}
\subsection{Construction}
The fermionic part of the Lagrangian density is
\be
{\cal L}_f = {\bar \psi}(i \gamma^\mu D_\mu -m ) \psi
\ee
with $ i D^\mu = i \partial^\mu - g A^\mu$. 

Moving to the light front coordinates, using the $A^{+}=0$ gauge and introducing 
the transverse lattice, 
\be
{\cal L}_f &=& {\psi^+}^\dagger ( i \partial^- - g A^-) \psi^+ +
{\psi^-}^\dagger i \partial^+ \psi^- \nonumber \\
&~&- i {\psi^-}^\dagger \alpha_r D^f_r \psi^+ - i {\psi^+}^\dagger \alpha_r
D^b_r \psi^- \nonumber \\
&~& - m {\psi^-}^\dagger \gamma^0 \psi^+ - m {\psi^+}^\dagger \gamma^0
\psi^-. \label{lfb}
\ee
 Here $r=1,2$ and $D^{f/b}_r $  is the forward/backward covariant
lattice derivative. Our goal here is to write the most local lattice
derivative.  That is why, instead of using the symmetric lattice
derivative,  in the above we have used the forward and backward
lattice derivatives. However,  
the Hermiticity of the Lagrangian (Hamiltonian)  requires that if one
of the  covariant lattice derivatives 
appearing in Eq. (\ref{lfb}) is  the forward derivative, the other has 
to be the backward derivative or vice versa.  
The covariant forward  and  backward derivatives on the lattice are
 defined as
\be
D_r^f \eta(\bx)= { 1 \over  a} [ U_r(\bx) \eta(\bx + a \br) - \eta(\bx)]
\ee
and
\be
D_r^b \eta(\bx) = { 1 \over a} [ \eta(\bx) - U_r^\dagger(\bx -a \br)
\eta(\bx -a \br)],
\ee
where $a$ is the lattice constant and $\br$ is unit vector in the
direction $r=1,2$ and ${D^f_r}^{\dagger} =-D^b_r$. $U_r(\bx)$ is the
group valued lattice gauge field with the property $U_r^{\dagger}(\bx) 
=U_{-r}(\bx+a\br)$.
Using the constraint equation
\be
i \partial^+ \psi^- = ( i \alpha_r D^f_r + \gamma^0 m) \psi^+,
\ee
and finally going over to the two component fields $\eta$
\be
{\cal L}_f &=& {\psi^+}^\dagger ( i \partial^- - g A^-) \psi^+ 
     - m {\psi^+}^\dagger \gamma^0 \psi^-  
- i {\psi^+}^\dagger
\alpha_r D^b_r \psi^-             \nonumber \\
&=&  {\psi^+}^\dagger ( i \partial^- - g A^-) \psi^+ \nonumber \\
&~&~~~~ - {\psi^+}^\dagger [ i \alpha_r D^b_r + \gamma^0 m ]{ 1 \over i
\partial^+} [ i \alpha_s D^f_s + \gamma^0 m] \psi^+ \nonumber \\
&=&  {\eta}^\dagger ( i \partial^- - g A^-) \eta \nonumber \\
&~&~~~~ - \eta^\dagger [ i {\hat \sigma}_r D^b_r -i m ]
{ 1 \over i \partial^+} [ i {\hat \sigma}_s D^f_s + i m] \eta ~. \label{lfb2}\nonumber \\ 
\ee  
$ {\hat \sigma_1} = \sigma_2$ and $ {\hat \sigma_2}= - \sigma_1$.
So, we arrive at the Lagrangian density 
\be {\cal L}_f &=& \eta^\dagger ( i \partial^- - g A^-) \eta
- m^2 \eta^\dagger { 1 \over i \partial^+}\eta \nonumber \\
& ~& ~~ -m \eta^\dagger(\bx) \sum_r {\hat
\sigma}_r {1 \over a}{1 \over  i \partial^+}\Big [
U_r(\bx) \eta(\bx + a \br) - \eta(\bx) \Big ] 
\nonumber \\
&~&~~- m \sum_r
\Big [
\eta^\dagger(\bx + a \br) U_r^\dagger(\bx) -
\eta^\dagger(\bx) \Big ] 
 {\hat \sigma}_r{ 1 \over a} { 1 \over 
i \partial^+} \eta(\bx) \nonumber \\
&~& ~- { 1 \over a^2}\sum_r [ \eta^\dagger(\bx + a \br) U_r^\dagger(\bx) -
\eta^\dagger(\bx)] {\hat \sigma}_r { 1 \over i \partial^+}{\hat
\sigma}_s [ U_s(\bx) \eta(\bx + a \bs) - \eta(\bx)].
\ee

In the free limit the fermionic part of the 
 Hamiltonian becomes 
\be
P^-_{fb} &=& \int dx^- a^2 \sum_{\bx} {\cal H} \nonumber \\
& = & \int dx^- a^2 \sum_{\bx} \Bigg [ m^2 \eta^\dagger(\bx) 
{ 1 \over i \partial^+}\eta(\bx) \nonumber \\
&~&~~~~ - { 1 \over a^2}  \eta^\dagger(\bx) \sum_r
{ 1 \over i \partial^+} [ \eta(\bx + a \br) - 2 \eta(\bx)
+ \eta(\bx - a \br) \nonumber \\
&~&~~~~ + { 1 \over a^2} \eta^\dagger(\bx)\sum_r (am {\hat \sigma_r}) 
{ 1 \over i \partial^+} [ \eta(\bx + a \br) - 2 \eta(\bx)
+ \eta(\bx - a \br) ] \Bigg ].\label{fb}
\ee
 In order to get Eq. (\ref{fb}), we have assumed infinite transverse lattice
and accordingly  have used shifting of lattice points.  
The positive  sign in front of the last term would change if we had
switched  forward and backward  
derivatives.

Because of the presence of ${\hat \sigma}_r$  the last term of Eq. (\ref{fb})  couples 
fermions of opposite helicities. Note that it is also linear in mass.
Such a helicity flip linear mass term is typical in continuum light
front QCD.  Here in free transverse lattice theory this 
 term arises from the interference of the first 
order derivative term and the mass term, due to the constraint equation. 
This is in contrast to the conventional lattice (see Appendix A) where no 
helicity flip or chirality-mixing term arises in the free theory if we use forward and 
backward lattice derivatives.   
 
In DLCQ the Hamiltonian is given by,
\be
H_{fb}= H_0 + H_{hf}
\ee
where
\be
H_0 &=& \sum_{\bz} \sum_\lambda \sum_l {a^2 m^2
\over l} \left [
b^\dagger(l,\bz,\lambda) b(l, \bz, \lambda)
+ d^\dagger(l,\bz,\lambda) d(l, \bz, \lambda) \right ] \nonumber \\
&~ &- \sum_{\bz} \sum_r \sum_\lambda \sum_{\lambda'}
\sum_l{ 1 \over l}
{\chi^\dagger}_{\lambda'}  \chi_\lambda 
\nonumber \\
&~& \Big [ b^\dagger(l, \bz, \lambda') b(l, \bz + a \br, \lambda)
- 2 b^\dagger(l, \bz, \lambda') b(l, \bz, \lambda)
+ b^{\dagger}(l, \bz, \lambda') b(l, \bz - a \br, \lambda)
\nonumber \\
&~&+d^\dagger(l, \bz, \lambda') d(l, \bz + a \br, \lambda)
- 2 d^\dagger(l, \bz, \lambda') d(l, \bz, \lambda)
+ d^{\dagger}(l, \bz, \lambda') d(l, \bz - a \br, \sigma)
\Big ] \label{fbnf}
\ee
and
\be
H_{hf} &= & \sum_{\bz} \sum_r \sum_\lambda \sum_{\lambda'}
\sum_l{ 1 \over l}
{\chi^\dagger}_{\lambda'} [   am {\hat \sigma}^r ] \chi_\lambda 
\nonumber \\
&~& \Big [ b^\dagger(l, \bz, \lambda') b(l, \bz + a \br, \lambda)
- 2 b^\dagger(l, \bz, \lambda') b(l, \bz, \lambda)
+ b^{\dagger}(l, \bz, \lambda') b(l, \bz - a \br, \lambda)
\nonumber \\
&~&+d^\dagger(l, \bz, \lambda') d(l, \bz + a \br, \lambda)
- 2 d^\dagger(l, \bz, \lambda') d(l, \bz, \lambda)
+ d^{\dagger}(l, \bz, \lambda') d(l, \bz - a \br, \sigma)
\Big ]. \label{fbf}
\nonumber \\
\ee

\subsection{Absence of doubling}

Consider the Fourier transform in transverse space
\be
\eta(x^-, \bx) = \int {d^2 k \over (2 \pi)^2} e^{i {\bf k} \cdot \bx}
\phi_{\bf k}(x^-)
\ee
where
$ - {\pi \over a} \le k_1, k_2 \le + {\pi \over a} $.
Then  the helicity nonflip part of Eq. (\ref{fb}) becomes
\be
 P^-_{nf} = \int dx^- \int { d^2 k \over (2 \pi)^2} \int { d^2 p \over (2
\pi)^2}
\phi^\dagger_{\bf k} (x^-){ 1 \over i \partial^+} \phi_{\bf p}(x^-) a^2
\sum_\bx  e^{-i ({\bf k} - {\bf p}) \cdot \bx} 
 \Bigg [ m^2 -\sum_r{ 1 \over a^2} \Big [ e^{ i {\bf p} \cdot a \br} - 2 + 
e^{-i{\bf p} \cdot a \br} \Big ] \Bigg ].
\ee
Using
\be
a^2 \sum_\bx e^{i ({\bf k} - {\bf p}) \cdot \bx}
= ( 2 \pi)^2 ~ \delta^2({\bf k} - {\bf p})
\ee
we get, 
\be
P^-_{nf} = \int dx^- \int { d^2 k \over (2 \pi)^2} \phi^\dagger_{\bf k}(x^-) { 1 \over
i \partial^+}
\phi_{\bf k}(x^-) \Bigg [ m^2 + \sum_r k_r^2  \left ({\sin~ k_ra/2 \over k_r
a/2}\right )^2 \Bigg ]
\ee
where we have defined $ k_ra = {\bf k} \cdot  \br a$.
Note that the $ sine $ function vanishes at the origin $k_1, k_2=0$
but does not vanish at the edges of the Brillouin zone $ k_1, k_2 =
\pm {\pi\over a} $. 

Define $ {\tilde k}_r =  k_r  {\sin ~ k_ra/2 \over k_r a/2}$. In the
naive continuum limit $ {\tilde k}_r \rightarrow k_r$.
  
Now, let us consider the  full Hamiltonian  (\ref{fb}) including the helicity flip term.
In the helicity space we have the following matrix structure for
$P^+P^-$ (since $P^-$ is inversely proportional to the total longitudinal
momentum $P^+$, we study the operator $P^+P^-$) 
\be
\begin{pmatrix} {m^2 + { 4 \over a^2} \sum_r \sin^2{k_ra \over 2} &
-{4m \over a} (i \sin^2{k_xa \over 2} + \sin^2 {k_ya \over 2} \cr
 & \cr
{4 m \over a} (i \sin^2 {k_x a \over 2} - \sin^2 {k_y a \over 2}) &
 m^2 + { 4 \over a^2} \sum_r \sin^2{k_ra \over 2}} 
\end{pmatrix}
\ee
which leads to the eigenvalue equation
\be
{\cal M}^2 = m^2 + {4 \over a^2} \sum_r \sin^2{k_r a \over 2} \pm {4 m \over
a} \sqrt{\sum_r \sin^4{k_r a \over 2}}.\label{eval}
\ee

Third term in the above equation comes from the linear mass helicity
flip term. If the mass $m=0$, then it is obvious from Eq. (\ref{eval}) 
that ${\cal M}^2=0$ if and only if $k_1=k_2=0$.  For nonzero $m$, one
can also in general conclude that ${\cal M}^2=m^2$ only for the case
$k_1=k_2=0$.  Thus there are no fermion doublers in this case (for
physical masses $am<1$). 
 In the following for  specific choices of momenta we elaborate on this
further.

If one component of the momentum vanishes, then 
\be
{\cal M}^2 = m^2 + {4 \over
a}({ 1 \over a}\pm m)\sin^2{ka \over 2} \label{fbe1}
\ee
 where $k$ is the non-vanishing
momentum component. Thus for $am=1$, irrespective of the value of $k$
we get ${\cal M}^2=m^2$ which is unwanted. In general, for $am >1$, 
 ${\cal M}^2$ can become negative. It is important to recall
that physical particles have $ m < { 1 \over a}$ (the lattice cutoff)
and hence are free from the species doubling on the lattice.
With periodic boundary condition (discussed in the next subsection),
 allowed $k$ values are $ k_q = \pm {
2 \pi q \over 2n+1}$, with $q=1,2,3, ..., n$ for $2n+1$ lattice sites
in each direction.  
Let $k_1=0$. For $ma=1.0$, Eq. (\ref{fbe1}) with 
the minus sign within the bracket gives ${\cal M}^2 =  m^2$ for all values
of $k_2$ and we get $2(2n+1)$-fold degenerate ground state with eigenvalue
$m^2.$\\ 
 The two spin states (spin up and down) are degenerate for
$k_1=k_2=0$. But if any one (or both) of the two 
transverse momenta is (are) nonzero then the
degeneracy is broken on the lattice by the spin flip term proportional to $m$.
So the total degeneracy of the lowest states for $ma=1.0 $ can 
be calculated in the
following  way:(a)
$~~k_1=k_2=0~~$: Number of states =2(spin up and spin down),
(b)$~~k_1=0,~~k_2\ne 0~~$: Number of states =2$n$ and
(c)$~~k_1\ne 0,~~k_2=0~~$: Number of states =2$n$.
Note that $k_i$ can have $2n$ nonzero
values and there is no spin degeneracy  for any nonzero $k_i$. 
So, the total number of degenerate states = $2+2n+2n = 2(2n+1).$
But if $ma \ne 1$ we cannot have $m^2$ eigenvalue for nonzero $k_i$ and we
have only two (spin) degenerate states with eigenvalue  $m^2$. 
Again we see from 
Eq. (\ref{fbe1}) that if $ma > 1$, the kinetic energy term becomes negative
and the eigenvalues go below $m^2$. But $ma \ge 1$ means $m \ge {1\over a}$
(ultra violet lattice cutoff) and hence unphysical.

\subsection{Numerical Investigation}
We have investigated the effects of two types of boundary conditions:
(1) fixed boundary condition and (2) periodic boundary condition.  
\subsubsection{Fixed boundary condition}
For each transverse direction, we choose $2n+1$ lattice points 
ranging from $-n$ to $+n$ where fermions are
allowed to hop. To implement fixed boundary condition we add two more
points at the two ends and demand that the fermion remains fixed at
these lattice points.  Thus we consider $2n+3$ lattice points.
Let us denote the fermion wavefunction at the location $s$ by $u(s)$. We have 
$ u(s) \sim \sin (s-1)ka$ with $u(1)=u(2n+3)=0$.  Allowed values of
$k$ are $(2n+2)k_pa=p \pi$
with $ p=1,2,3, ...., 2n+1$ and $ k_p= {\pi \over (2n+2)a}p$. Thus the
minimum $k_p$ allowed is ${ \pi  \over a}{ 1 \over (2n+2)}$ and maximum $k_p$
allowed is ${\pi \over a} {(2n+1) \over (2n+2)}$. For example, for $n=1$
we have   $k_1={\pi \over 4a},~ k_2={2 \pi \over 4 a},~ k_3 = {3 \pi
\over 4 a}$, etc. 
\subsubsection{Periodic boundary condition}

Again, for each transverse direction, we choose $2n+1$ lattice points. 
We identify the
$(2n+2)^{th}$ lattice point with the first lattice point. 
In this case we have the  fermion wavefunction $ u(s) \sim e^{iska}$ with the
condition $u(s) = u(s+L)$ where $L=2n+1$. Thus $ (2n+1)k_p a = \pm 2
\pi p$ so that $k_p = \pm { 2 \pi \over (2n+1)}p$, $p=0,1,2, ...,
n$. Thus the minimum $k_p$ allowed is $0$ and the maximum $k_p$
allowed is $ \pi {2 n \over 2n+1}$.  
For $n=1$, we have, $ k_0=0, k_1=\pm {2 \over 3} \pi$, etc.
\subsubsection{Numerical results}
For the study of the fermion spectra on the transverse lattice, the
longitudinal momentum plays a passive role and for the numerical
studies we choose the dimensionless longitudinal momentum ($l$) to be unity
which is kept fixed.  For a given set of lattice points in the
transverse space we diagonalize the Hamiltonian and compute both
eigenvalues and eigenfunctions. 

First we discuss the results for $H_0$ given in Eq. (\ref{fbnf}).
We  diagonalize the Hamiltonian using basis states defined at each
lattice point in a finite region in the transverse plane. Let us
denote a general lattice point in the transverse plane by
$(x_i,y_i)$.  For each choice of $n$ (measure of the linear lattice size), we have 
$ -n \le x_i,y_i \le +n $. Thus for a given $n$, we have a $(2n+1) \times (2n+1)$
dimensional matrix for the Hamiltonian. 
The boundary conditions do have significant effects at small
volumes. For example, a zero transverse momentum fermion at {\it finite}
$n$ is/not allowed with periodic/fixed boundary condition.
With fixed boundary condition, in
the infinite volume limit, we expect the lowest eigenstate to be the
zero transverse momentum fermion with the eigenvalue $m^2$. In Fig. 1 we
show the convergence of the lowest eigenvalue as a function of $n$
towards the infinite volume limit in this case ($m=1$ in Fig. 1).  

For a zero transverse momentum fermion, the probability amplitude to
be at any transverse location should be independent of the transverse
location. Thus we expect the eigenfunction for such a particle to be a
constant. At finite volume, with fixed boundary condition, we do get a 
nodeless wave function which
nevertheless is not a constant since it carries some non-zero
transverse momentum. All the excited states carry non-zero transverse
momentum in the infinite volume limit. All of them have nodes
characteristic of sine waves. The eigenfunctions corresponding to the
first three eigenvalues are shown in Fig. 2 for the case of fixed
boundary condition.
With periodic boundary condition, for any $n$, we get a zero
transverse momentum fermion with a flat wave function.

Now, we consider the effect of helicity flip term. 
With fixed boundary condition the lowest eigenstate has non vanishing 
transverse momentum in finite volume. In the absence of helicity flip term
positive and negative helicity fermions are degenerate. The helicity flip 
term  lifts the degeneracy. The splitting is larger for larger transverse
momentum. 
In Fig. 3 we present the level splitting  for the helicity up and down 
fermions as a
function of $n$. As expected, the level splitting vanishes and we get
exact degeneracy in the infinite volume limit. 
For the periodic boundary condition, the lowest state has exactly zero
transverse momentum and we get two degenerate fermions for all $n$.
\section{Hamiltonian with symmetric derivative}
\subsection{construction}
The symmetric derivative is defined by
\be
D_r \psi^\pm(\bx) = { 1 \over 2 a} [ U_r(\bx) \psi^\pm(\bx + a \br) - 
U_{-r}(\bx) \psi^\pm(\bx - a \br)].
\ee
In place of using forward and backward derivatives in Eq. (\ref{lfb}), 
we use the above symmetric derivative for all lattice
derivatives. Proceeding as in Sec. IIIA, we arrive at the 
 fermionic part of the QCD Hamiltonian 
\be
P^-_{sd}&= &\int dx^- a^2 \sum_\bx
 m^2
\eta^\dagger(\bx) { 1 \over i \partial^+} \eta( \bx) \nonumber \\
 & ~ &  -\int dx^- a^2 \sum_\bx
\Bigg \{  m \frac{1}{2a}
\eta^\dagger(\bx) \sum_r {\hat \sigma}_r { 1 \over i \partial^+}  
\left [ U_{r}(\bx) \eta(\bx + a \br) - U_{-r}(\bx) \eta(\bx - a \br) \right ] 
\nonumber \\
& ~ & -  m  \frac{1}{2a}
 \sum_r \left [ \eta^\dagger(\bx - a \br) {\hat \sigma}_r U_{r}(\bx - a \br) - 
 \eta^\dagger(\bx + a \br) {\hat \sigma}_r U_{-r}(\bx + a \br)  \right ]
{ 1 \over i \partial^+} \eta(\bx) \Bigg \}  \nonumber \\
 & ~ & -\int dx^- a^2 \sum_\bx
 \frac{1}{4a^2} \sum_r 
\Big [ \eta^\dagger(\bx - a \br)  U_{r}(\bx - a \br) - 
\eta^\dagger(\bx + a \br)   U_{-r}(\bx + a \br) \Big ]
\nonumber \\
&~&~~~~~{ 1 \over i \partial^+} 
\Big [  U_{r}(\bx) \eta(\bx + a \br) - U_{-r}(\bx) \eta(\bx - a \br)\Big ] .
\label{qcdsd}
\ee 
In the free limit, the above Hamiltonian becomes
\be
P^-_{sd} &=&\int dx^- a^2 \sum_\bx \Bigg\{
 m^2 {\eta}^\dagger(\bx) { 1 \over i \partial^+}
\eta(\bx) \nonumber \\
&~& + { 1 \over 4 a^2} \sum_r [{\eta}^\dagger(\bx + a \br) - 
{\eta}^\dagger(\bx - a \br)]{ 1 \over i \partial^+}[\eta(\bx + a
\br) - \eta(\bx - a \br)]\Bigg\}.\label{sd}
\ee
In the free field limit the two linear mass terms cancel with each
other.

Using DLCQ for the longitudinal direction, we get 
\be
P_{sd}^- ={ L \over \pi} H_{sd} \equiv { L \over \pi} [ H_m + H_k]
\ee
with 
\be
H_m  &= & a^2{m}^2 \sum_l \sum_\sigma \sum_\bz { 1 \over l} \nonumber \\
&~&~~~~ [ b^\dagger(l,\bz, \sigma) b(l,\bz, \sigma) + d^\dagger(l,\bz, \sigma)
d(l,\bz, \sigma) ]
\ee
and 
\be
H_k & = & \sum_l \sum_\sigma \sum_\bz \sum_r {1 \over l}\nonumber \\
&~& ~~~ \Bigg [ 
b^\dagger(l,\bz+ a \br, \sigma) b(l,\bz+a \br, \sigma) +
b^\dagger(l,\bz- a \br, \sigma) b(l,\bz- a \br, \sigma) \nonumber \\
&~&~~~- b^\dagger(l,\bz+ a \br, \sigma) b(l,\bz-a \br, \sigma)
-b^\dagger(l,\bz-a \br, \sigma) b(l,\bz+ a \br, \sigma) \nonumber \\
&~&~~~+d^\dagger(l,\bz+ a \br, \sigma) d(l,\bz+a \br, \sigma) +
d^\dagger(l,\bz- a \br, \sigma) d(l,\bz- a \br, \sigma) \nonumber \\
&~&~~~- d^\dagger(l,\bz+ a \br, \sigma) d(l,\bz-a \br, \sigma)
-d^\dagger(l,\bz-a \br, \sigma) d(l,\bz+ a \br, \sigma) \Bigg ].
\ee

When we implement the constraint equation on the lattice and use
symmetric definition of the lattice derivative, it is important to
keep in mind that we have only next to nearest neighbor
interactions. Thus a decoupling of even and odd lattice points occur
and as a result we have two independent sub-lattices one connecting odd
lattice points and the other connecting even lattice points. 

Let us now address the nature of the spectrum and the presence
of doublers. 

\subsection{Fermion doubling}
 The Hamiltonian (\ref{sd}) can be  rewritten as
\be
 P^-_{sd}&=& \int dx^- a^2\sum_{\bx=even}\Big [
m^2 \eta^{\dagger}(\bx){1 \over i\partial^+}\eta(\bx)
-{1 \over 4 a^2} a^2  [ \eta^\dagger(\bx) { 1 \over i \partial^+}\sum_r [
\eta(\bx +2 a \br) ] + \eta(\bx -2 a \br)- 2\eta(\bx )]\Big ]
\nonumber \\
&~&+ \int dx^- a^2\sum_{\bx=odd}\Big [
m^2 \eta^{\dagger}(\bx){1 \over i\partial^+}\eta(\bx)
-{1 \over 4 a^2} a^2  [ \eta^\dagger(\bx) { 1 \over i \partial^+}\sum_r [
\eta(\bx +2 a \br) ] + \eta(\bx -2 a \br)- 2\eta(\bx )]\Big ].
\ee
Clearly the Hamiltonian is divided into even and odd sub-lattices each
with lattice constant $2a$. As a result, a momentum component in each
sub-lattice is bounded by  ${\pi \over 2a}$ in magnitude. 
Again, going through the Fourier transform in each sub-lattice of the
transverse space, we arrive at
 the free particle dispersion relation for the light
front energy in each sector
\be
k^-_{\bf k} = { 1 \over k^+} [ m^2 + { 1 \over a^2}
\sum_r \sin^2~ k_ra ]. \label{potdoub}
\ee
For fixed $k_r$, in the limit $ a \rightarrow 0$ $ { 1 \over a^2}
\sin^2~k_ra \rightarrow k_r^2$ and we get the continuum dispersion
relation
\be
k^-_{\bf k} = { m^2 + {\bf k}^2 \over k^+}.
\ee
        
  Because of the momentum bound of ${\pi \over 2a}$
 doublers cannot arise from
$ka=\pi$. However, because of the decoupling of odd and even lattices,
one can get two zero transverse momentum fermions one each from the
two sub-lattices. Thus, for two transverse dimensions, we can get four
zero transverse momentum fermions as follows: (1) even lattice points in
$x$, even lattice points in $y$, (2) even lattice points in $x$, odd
lattice points in $y$, (3) odd lattice points in $x$, even lattice
points in $y$, and (4) odd lattice points in $x$, odd lattice points
in $y$. Thus we expect a four fold degeneracy of zero transverse
momentum fermions. 

\subsection{Numerical Investigation}
\subsubsection{Fixed boundary condition}
For each transverse direction,  we have $2n+1$ lattice points where the 
fermions are
allowed to hop. To implement the fixed boundary condition, we need to 
consider $2n+5$ lattice points. For one sub-lattice we have to fix
particles at $s=1$ and $s=2n+5$. We have, the wavefunction at location
$s$, $u_s \sim \sin~(s-1)ka$. We have, $u_s=0$ for $ s=1$. We also
need $u_s=0$ for $s=2n+5$. Thus $ (2n+4)k_pa =p \pi$, with $p=1,2,3,
..., n+1$. For $n=1$, allowed values of $k_p$ are $k_p={ \pi \over
6a}, { 2\pi \over 6a}$.

For the other sub-lattice, we fix the particles at $s=2$ and
$s=2n+4$. The wavefunction at location $s$, $u_s \sim
\sin~(s-2)ka$. $u_s=0$ for $s=2$ and $s=2n+4$. Thus $ (2n+2)k_pa = p
\pi$ 
with $ p = 1,2,3, ...., n$. For $n=1$, only allowed value
of $k$ is $k={\pi \over 4a}$.

Combining the two sub-lattices, for $n=1$, the allowed values of $k$ are
${\pi \over 6a}, {\pi \over 4a},$ and $ {2 \pi \over 6a}$. 
\subsubsection{Periodic boundary condition}

For a given $n$, fermions are allowed to hop at $2n+1$ lattice
points in each transverse directions. Consider $2n+3$ lattice points. 
For one sub-lattice $(2n+3)^{rd}$ lattice
point is identified with the lattice point 1. For the other sub-lattice
$(2n+2)^{nd}$ lattice point is identified with the lattice point
2. Wavefunction at point $s$, $u_s \sim e^{is ka}$. We require $
e^{ika} =e^{ i (2n+3)ka}$. Thus $k_pa = \pm { 2 \pi p \over
(2n+2)}$, $ p=0,1,2, ..., {n+1 \over 2}$.  For $n=1$, we have, $k_0=0, k_1=\pm
{\pi \over 2a}$.

For the other sub-lattice we require $e^{ 2ika} = e^{i(2n+2)ka}$. Thus
$ k_pa = \pm {\pi \over n}p$, $p = 0,1,2, ..., {n-1 \over 2}$. 
For $n=1$, allowed
value of $k=0$. Thus for $n=1$, taking the two
sub-lattices together, the allowed values of $k$ are $0,0, {\pi \over
2a}$.   

\subsubsection{Numerical results}
The results of matrix diagonalization in the case of the symmetric derivative 
with fixed boundary condition are presented in Figs. 4, 5 and 6. 
In Fig. 4 we present the lowest four
eigenvalues as a function of $n$. At finite volume, the four states do
not appear exactly degenerate even though the even-odd and odd-even
states are always degenerate because of the hypercubic (square) symmetry 
in the transverse
plane. The four states become degenerate in the infinite volume
limit. The eigenfunctions of the lowest four states are presented in
Fig. 5 for $n=5$. As they correspond to particle states, they are
nodeless. All other states in the spectrum have one or more nodes. For
example, in Fig. 6 we show the eigenfunction corresponding to the
fifth eigenvalue which clearly exhibits the node structure.

With periodic boundary condition, for any $n$ we get four degenerate
eigenvalues corresponding to zero transverse momentum fermions. 
Corresponding wavefunctions are flat in transverse coordinate space.

\section{Staggered fermion on the light front transverse lattice}
\label{lfsf}
As we have seen in the previous section that the method of symmetric
derivatives results in fermion doublers, we now consider 
 two approaches to  remove the doublers. In this section we study an
approach similar to the staggered fermions in conventional lattice
gauge theory. In the next section we will take up the case of
 Wilson fermions.

In analogy with the Euclidean staggered formulation, define the spin 
diagonalization transformation
\be
\eta(x_1,x_2) = ({\hat\sigma}^1)^{x_1}({\hat\sigma}^2)^{x_2}
\chi(x_1,x_2). \label{sdt}
\ee
We see from the  QCD Hamiltonian given in Eq. (\ref{qcdsd}) with symmetric 
derivative that  in the interacting 
theory (except  for the linear mass term) and also  in the free fermion limit,
even and odd lattice sites are decoupled and the Hamiltonian 
 is already spin diagonal. So, it is very natural to try staggered fermion 
formulation on the light front transverse lattice.  In this section we shall follow 
the Kogut-Susskind formulation \cite{ks} and  present an elementary
configuration space analysis for two flavor interpretation.
After the spin transformation  the  linear mass term in the
Hamiltonian (\ref{qcdsd})  becomes:
\be
&~&\int dx^- a^2 \sum_\bx \Bigg \{  m \frac{1}{2a}
\chi^\dagger(\bx) \sum_r \phi(\bx,r) { 1 \over i \partial^+}  
\left [ U_{r}(\bx) \chi(\bx + a \br) - U_{-r}(\bx) \chi(\bx - a \br) \right ] 
\nonumber \\
& ~ & -  m  \frac{1}{2a}
 \sum_r \left [ \chi^\dagger(\bx - a \br) \phi(\bx,r)U_{r}(\bx - a \br) - 
 \chi^\dagger(\bx + a \br) \phi(\bx,r) U_{-r}(\bx + a \br)  \right ]
{ 1 \over i \partial^+} \chi(\bx) \Bigg \}  
\ee
where, $\phi(\bx,r) = 1$ for $r=1$ and $\phi(\bx,r) = (-1)^{x_1}$ for $r=2$.
After spin diagonalization, the  full Hamiltonian in the free field limit becomes 
\be
  P^-_{sf}&=&\int dx^- a^2 \sum_\bx \Bigg\{ 
 m^2 {\chi}^\dagger(\bx) { 1 \over i \partial^+}
\chi(\bx) \nonumber \\
&~& + { 1 \over 4 a^2} \sum_r [{\chi}^\dagger(\bx + a \br) - 
{\chi}^\dagger(\bx - a \br)]{ 1 \over i \partial^+}[\chi(\bx + a
\br) - \chi(\bx - a \br)] \nonumber \\
&~& -{1\over 2a}m {\chi}^\dagger(\bx){ 1 \over i \partial^+} 
\sum_r \phi(\bx,r)
[\chi(\bx + a\br) - \chi(\bx - a \br)] \nonumber \\
&~& -{1\over 2a}m\sum_r [{\chi}^\dagger(\bx + a \br) - {\chi}^\dagger
(\bx - a \br)]
\phi(\bx,r){ 1 \over i \partial^+}\chi(\bx)
\Bigg\}.\label{stag}
\ee  
 The two linear mass terms cancel with each other in the free theory,
but since they are present
in the interacting theory we keep them to investigate the staggered
fermions. 

Since all the terms in Eq. (\ref{stag}) are spin diagonal,
 we can put only a single component field at each transverse site.
From now on, all the $\chi$'s and $\chi^{\dagger}$'s appearing in
Eq. (\ref{stag}) can be taken as single component fermion fields. 
Thus we have thinned the fermionic degrees of freedom by half.
Without loss of generality, we keep the helicity up component of $\chi$ at 
each lattice point.

Apart from the linear mass term in Eq. (\ref{stag}), all the other
terms have the feature that fermion fields on the even and odd
lattices do not mix. Let us denote (see Fig. 7) the even-even lattice points by 1,
odd-odd lattice points by $1'$, odd-even lattice points by 2 and
even-odd lattice points by $2'$, and the corresponding fields by
$\chi_1$ etc. Then  the first of the linear mass terms    
\be
  \sum_{\bx}{\chi}^\dagger(\bx){ 1 \over i \partial^+} \sum_r \phi(\bx,r)
[\chi(\bx + a\br) - \chi(\bx - a \br)]
\ee
can be rewritten as (suppressing  factors of $a$),  
\be
 &~& {\chi_1}^\dagger{ 1 \over i \partial^+}( \nabla_1\chi_2+\nabla_2
\chi_{2^{\prime}})
+{\chi_2}^\dagger{ 1 \over i \partial^+}( \nabla_1\chi_1-\nabla_2\chi_
{1^{\prime} })\nonumber \\
&~&+{\chi_{1^{\prime }}}^\dagger{ 1 \over i \partial^+}( \nabla_1
\chi_{2^{\prime} } -
\nabla_2\chi_2)
+{\chi_{2^{\prime }}}^\dagger{ 1 \over i \partial^+}( \nabla_1\chi_{1^{\prime }}+
\nabla_2\chi_1) +B \label{sc}
\ee
where $\nabla_1$ and $\nabla_2$ are the symmetric derivatives in
the respective directions. Looking at Fig. 7 it is apparent that these 
$\nabla_1$ and $\nabla_2$ can also be interpreted as a block
derivative, i.e., finite differences between block variables. For example,
$\nabla_1\chi_1 =\chi_1(1,0)-\chi_1(0,0)$.  $B$ represents the
contribution from other blocks.

Using Eq. (\ref{sdt}),  in terms of the nonvanishing components of $\eta$, we have
\be
\eta_1=\chi_1,~~\eta_2=i\chi_2,~~\eta_{ 1'}=i\chi_{ 1'},~~
\eta_{2' }=-\chi_{ 2'}.
\ee
An interesting feature of lattice points 1 and $1'$ is that fermion
fields $\eta_1$ and $\eta_{1'}$ have positive helicity. $\eta_2$ and
$\eta_{2'}$ have negative helicity.
In terms of $\eta$ fields the expression given in Eq. (\ref{sc})  can be written as
\be
&~& {\eta_1}^\dagger{ 1 \over i \partial^+}(-i \nabla_1\eta_2 -\nabla_2
\eta_{2^{\prime }})
+i{\eta_2}^\dagger{ 1 \over i \partial^+}( \nabla_1\eta_1+i\nabla_2\eta_{1^{
\prime }}) \nonumber \\
&~&+i{\eta_{1^{\prime }}}^\dagger{ 1 \over i \partial^+}(- \nabla_1
\eta_{2^{\prime} }+i\nabla_2\eta_2)
-{\eta_{2^{\prime}}}^\dagger{ 1 \over i \partial^+}(-i \nabla_1\eta_{1^{\prime }}+
\nabla_2\eta_1) +B. \label{stag2}
\ee
Now,  
\be
\eta(1)-\eta(0)&=& {1\over 2}(\eta(1)-\eta(-1)) +{1\over 2}(\eta(1)+\eta(-1) 
-2\eta(0)) \nonumber \\
&\equiv& {\hat\nabla}\eta(0) +{1\over 2} {\hat\nabla}^2\eta(0)~,
\ee
\be
\eta(0)-\eta(-1)&=& {1\over 2}(\eta(1)-\eta(-1)) -{1\over 2}(\eta(1)+\eta(-1) 
-2\eta(0)) \nonumber \\
&\equiv& {\hat\nabla}\eta(0) -{1\over 2} {\hat\nabla}^2\eta(0)
\ee
where $ {\hat\nabla} $ and  $ {\hat\nabla}^2 $ are respectively first order and 
second order block derivatives.
So, we can write the expression (\ref{stag2})  as
\be
&~& {\eta_1}^\dagger{ 1 \over i \partial^+}\Big \{-i({\hat \nabla}_1\eta_2 
-{1\over 2}{{\hat \nabla}_1}^2\eta_2)
-({\hat\nabla}_2\eta_{2^{\prime }} -{1\over 2}{{\hat\nabla}_2}^2
\eta_{2^{\prime }} \Big \} \nonumber \\
 & ~&+i{\eta_2}^\dagger{ 1 \over i \partial^+}\Big\{ ( {\hat\nabla}_1
\eta_1+{1\over 2} {{\hat\nabla}_1}^2\eta_1)
+i({\hat\nabla}_2\eta_{1^{\prime }} -{1\over 2}{{\hat\nabla}_2}^2
\eta_{1^{\prime }}) \Big \} \nonumber \\
 &~&+i{\eta_{1^{\prime }}}^\dagger{ 1 \over i \partial^+}\Big\{-( 
{\hat\nabla}_1\eta_{2^{\prime }}+{1\over 2}
{{\hat\nabla}_1}^2\eta_{2^{\prime }})
+i({\hat\nabla}_2\eta_2+{1\over 2}
{{\hat\nabla}_2}^2\eta_ 2 )\Big \}  \nonumber \\
&~&-{\eta_{2^{\prime }}}^\dagger{ 1 \over i \partial^+}\Big\{-i ({\hat\nabla}_1
\eta_{1^{\prime }}-{1\over 2}{{\hat\nabla}_1}^2\eta_{1^{\prime}})
+( {\hat\nabla}_2\eta_1 + {1\over 2} {{\hat\nabla}_2}^2\eta_1)\Big\}~.
\label{stag3} \ee
Let us introduce the fields
\be
u_1&=&{1\over {\sqrt 2}}(\eta_1+\eta_{1^{\prime}}) \nonumber \\
u_2&=&{1\over {\sqrt 2}}(\eta_2+\eta_{2^{\prime}}) \nonumber \\
\tilde{d}_1&=&{1\over {\sqrt 2}}(\eta_1-\eta_{1^{\prime}}) \nonumber \\
\tilde{d}_2&=&{1\over {\sqrt 2}}(\eta_2-\eta_{2^{\prime}}). \nonumber \\
\ee
Then, the first order derivative term in Eq. (\ref{stag3}) can be written as
\be
 u^{\dagger}{1\over i \partial^+}{\hat\sigma}^r{\hat\nabla}_r u +
d^{\dagger}{1\over i \partial^+}{\hat\sigma}^r{\hat\nabla}_r d 
= f^{\dagger}{1\over i \partial^+}{\hat\sigma}^r{\hat\nabla}_r f
\ee
where, $d={\hat\sigma}^1\tilde{d}$ and  
 the flavor isospin  doublet
\be
f = \left[ \begin{array}{l} u \\ d \end{array} \right] .
\ee
Similarly, we can write the second order block derivative term in
expression (\ref{stag3}) as
\be
{1\over 2}f^{\dagger}{1\over i \partial^+}\sigma^3{T}^r
{{\hat\nabla}_r}^2f
\ee
where, $T^r$s are the matrices in the flavor space defined as
\be
T^1 = -i\sigma^2,~~~
T^2 =  -i\sigma^1.
\ee
Similarly, the second term in Eq. (\ref{stag}) 
\be
 \sum_r [{\eta}^\dagger(\bx + a \br) - 
{\eta}^\dagger(\bx - a \br)]{ 1 \over i \partial^+}[\eta(\bx + a
\br) - \eta(\bx - a \br)]
\ee
reads as
\be
{\hat\nabla}_r f^{\dagger}{1\over i\partial^+}{\hat \nabla}_r f 
+{{\hat\nabla}_r}^2f^{\dagger}{1\over i \partial^+}{{\hat\nabla}_r}^2f 
+{i\over2}[{\hat\nabla}_rf^{\dagger}{1\over i \partial^+}\sigma^rT^r{{\hat
\nabla}_r}^2f +{{\hat\nabla}_r}^2f^{\dagger}{1\over i
\partial^+}\sigma^rT^r{{\hat  
\nabla}_r}f]~.
\ee
 The full Hamiltonian given in Eq.  (\ref{stag}) can 
now be written in two flavor notation as
\be
P^-_{sf}=\int dx^- a^2\sum_x &\Big\{& 
 m^2f^{\dagger}{1\over i \partial^+}f
 +{1\over 4 }[
{\hat\nabla}_r f^{\dagger}{1\over i\partial^+}{\hat \nabla}_r f    
+a^2{{\hat\nabla}_r}^2f^{\dagger}{1\over i \partial^+}{{\hat\nabla}_r}^2f
\nonumber \\
&~& +{ia\over2}({\hat\nabla}_rf^{\dagger}{1\over i
\partial^+}\sigma^rT^r{{\hat
\nabla}_r}^2f +{{\hat\nabla}_r}^2f^{\dagger}{1\over i
\partial^+}\sigma^rT^r{{\hat
\nabla}_r}f)]\nonumber \\
&~&-{1\over 2}m (f^{\dagger}{1\over i
\partial^+}{\hat\sigma}^r{\hat\nabla}_r
 f + {a\over 2}f^{\dagger}{1\over i \partial^+}\sigma^3{T}^r
{{\hat\nabla}_r}^2f + h.c )\Big \}~.\label{flavor}
 \ee
The above simple exercise shows that applying the spin diagonalization 
on the symmetric derivative method, the number of doublers on the
transverse lattice can be reduced from four to two which can be
reinterpreted as two flavors.   Although in the free case given by
Eq. (\ref{flavor}) the second and third lines are separately zero
identically, we have kept these terms because in QCD similar terms
will survive.  These terms exhibit flavor mixing and also helicity
flipping. The flavor mixing terms are always irrelevant.

\section{Wilson term on the light front transverse lattice}
Since doublers in the light front transverse lattice arise from the
decoupling of even and odd lattice sites, a term that
will couple these sites will remove the zero momentum doublers. However, 
 conventional doublers now may arise from the edges of the Brillouin zone.
 A second derivative term couples the even and odd lattice sites and also 
removes the conventional doublers. Thus,   
the term originally proposed by Wilson to
remove the doublers arising from $ka=\pi$ in the conventional lattice
theory will do the job \cite{BK}. 

To remove doublers, add an irrelevant term to the Lagrangian density
\be
\delta {\cal L}(\bx) = {\kappa \over a} \sum_{r}{\bar \psi}(\bx) [U_r(\bx) \psi(\bx 
+ a\br) - 2 \psi(\bx)+ U_{-r}(\bx)\psi(\bx - a \br)]
\ee
where $\kappa$ is the Wilson parameter.
This generates the following additional terms in the Hamiltonian 
 (\ref{qcdsd}):
\be
P^-_{w} & = &  -\int dx^- a^2 \sum_\bx
\Bigg \{  4 \frac{\kappa}{a} \frac{1}{2a}
\eta^\dagger(\bx) \sum_r {\hat \sigma}_r { 1 \over i \partial^+}  
\left [ U_{r}(\bx) \eta(\bx + a \br) - U_{-r}(\bx) \eta(\bx - a \br) \right ] 
\nonumber \\
& ~ & - 4 \frac{\kappa}{a}  \frac{1}{2a}
 \sum_r \left [ \eta^\dagger(\bx - a \br) {\hat \sigma}_r U_{r}(\bx - a \br) - 
 \eta^\dagger(\bx + a \br) {\hat \sigma}_r U_{-r}(\bx + a \br)  \right ]
{ 1 \over i \partial^+} \eta(\bx) \Bigg \}  \nonumber \\
 & ~ &+ \int dx^- a^2 \sum_\bx \Bigg \{
 \frac{\kappa}{a} \frac{1}{2a}\sum_r \sum_s 
\Big [ \eta^\dagger(\bx - a \br) U_{r}(\bx - a \br) + 
\eta^\dagger(\bx + a \br) U_{-r}(\bx + a \br) \Big ] \nonumber \\
&~&~~~~~{ 1 \over i \partial^+} {\hat \sigma}_s 
\Big [  U_{s}(\bx) \eta(\bx + a \bs) - U_{-s}(\bx) \eta(\bx - a \bs) \Big ] \nonumber \\
& ~ & - \frac{\kappa}{a} \frac{1}{2a}\sum_r \sum_s 
\Big [ \eta^\dagger(\bx - a \br) {\hat \sigma}_r U_{r}(\bx - a \br) - 
\eta^\dagger(\bx + a \br) {\hat \sigma}_r U_{-r}(\bx + a \br) \Big ]
\nonumber \\
&~&~~~~~{ 1 \over i \partial^+}  
\Big [  U_{s}(\bx) \eta(\bx + a \bs) + U_{-s}(\bx) \eta(\bx - a \bs)\Big ]
\Bigg \} \nonumber \\
& ~ & -\int dx^- a^2 \sum_\bx
\Bigg \{ \mu \frac{\kappa}{a} 
\eta^\dagger(\bx) { 1 \over i \partial^+}\sum_r    
\left [ U_{r}(\bx) \eta(\bx + a \br) 
+ U_{-r}(\bx) \eta(\bx - a \br) \right ] \nonumber \\
& ~ & +  \mu\frac{\kappa}{a}  
 \sum_r \left [ \eta^\dagger(\bx - a \br) U_{r}(\bx - a \br) + 
 \eta^\dagger(\bx + a \br) U_{-r}(\bx + a \br)  \right ]
{ 1 \over i \partial^+} \eta(\bx) \Bigg \} \nonumber \\
 & ~ & - \int dx^- a^2 \sum_\bx
  \frac{\kappa^2}{a^2} \sum_r \sum_s 
\Big [ \eta^\dagger(\bx - a \br)  U_{r}(\bx - a \br) + 
\eta^\dagger(\bx + a \br)  U_{-r}(\bx + a \br) \Big ] \nonumber \\
&~&~~~~~{ 1 \over i \partial^+}  
\Big [  U_{s}(\bx) \eta(\bx +a \bs)  + U_{-s}(\bx) \eta(\bx - a \bs)\Big ].
\ee
In addition, the factor $m^2$ in the free term in (\ref{qcdsd}) gets replaced by
$\mu^2=(m+4{\kappa \over a})^2$.

In the free limit the resulting Hamiltonian  goes over to
 
\be
 P^-_w &=& \int dx^- a^2 \sum_{\bx} \Bigg [ 
\mu^2 {\eta}^\dagger (\bx) { 1 \over i \partial^+} \eta(\bx)
\nonumber \\
&~& +
{ 1 \over 2 a} \sum_{r}[{\eta}^\dagger(\bx + a \br) - {\eta}^\dagger(\bx
-a \br)]{ 1 \over i \partial^+} {1 \over 2 a}[\eta(\bx + a \br) -
\eta(\bx - a \br)]
\nonumber \\
&~& + { \kappa^2 \over a^2}\sum_{r} [{\eta}^\dagger (\bx + a \br)
-2 {\eta}^\dagger(\bx) + {\eta}^\dagger(\bx -a \br)]
{ 1 \over i \partial^+}[\eta(\bx + a \br) - 2 \eta(\bx) +
\eta(\bx - a \br) ] \nonumber \\
&~& - 2 {\mu\kappa \over a}\sum_{r} {\eta}^\dagger(\bx) {1 \over i \partial^+}[
\eta(\bx + a \br) - 2 \eta(\bx) + \eta(\bx - a \br)] \Bigg ] . \label{w}
\ee
We  rewrite the free Hamiltonian (\ref{w}) as
\be
P_w^-=P_D^- + P_{OD1}^- +P_{OD2}^-.
\ee

The diagonal terms are
\be
P^-_D &=& \int dx^- a^2 \sum_{\bx} {\eta}^\dagger (\bx) { 1 \over i
\partial^+} \eta(\bx) \Big [ \mu^2 + { 1 \over a^2} + 8 \mu \kappa { 1 \over a} +
12 \kappa^2 { 1 \over a^2} \Big ].
\ee
The nearest neighbor interaction is
\be
P^-_{OD1} &=& -\int dx^- a^2 \sum_{\bx} \sum_{\br} \nonumber \\
&~& 
\Bigg [  (2 \mu\kappa { 1 \over a} + 4 {\kappa^2 \over a^2}) \Big [
{\eta}^\dagger (\bx) { 1 \over i \partial^+} \eta (\bx + a \br)
+ {\eta}^\dagger (\bx) { 1 \over i \partial^+} \eta (\bx + a \br) \Big ]
\Bigg ].
\ee
The next to nearest neighbor interaction is
\be P^-_{OD2} &=& \int dx^- a^2 \sum_{\bx} \sum_{\br} 
\Big \{ - { 1 \over 4 a^2} + { \kappa^2 \over a^2} \Big \}
\nonumber \\ 
&~& \Big [
{\eta}^\dagger(\bx + a \br) { 1 \over i \partial^+} \eta(\bx - a \br) +
{\eta}^\dagger(\bx - a \br) { 1 \over i \partial^+} \eta(\bx + a \br) 
\Big ]. \nonumber \\
\ee 

Using the Fourier transform in the transverse space, we get,
\be 
P^-_w&=&\int dx^- \int {d^2 k \over (2 \pi)^2)} \phi_{\bf k}^\dagger(x^-) {1 \over i
\partial^+} \phi_{\bf k}(x^-)  
\Bigg [ \mu^2 + \sum_r k_r^2 \Bigg ( {\sin k_ra \over k_r a} \Bigg
 )^2  \nonumber \\
&~&+ 2 a \mu \kappa \sum_r k_r^2 \Bigg ( {\sin k_ra/2 \over k_r a/2} \Bigg 
 )^2  
~+ a^2 \kappa^2  \sum_r k_r^4 \Bigg ( {\sin k_ra/2 \over k_r a/2} \Bigg 
 )^4 ~\Bigg ].
\ee
Note that, as anticipated, Wilson term removes the doublers because
the lowest eigenvalue occurs only if all the $k_r$'s are zero. 
       
In DLCQ, we have,
\be
H_D &= & [ a^2{\mu}^2 + 1+ 8 a \mu \kappa + 12 \kappa^2] \sum_l 
\sum_\sigma \sum_\bz { 1 \over l} \nonumber \\
&~&~~~~ [ b^\dagger(l,\bz, \sigma) b(l,\bz, \sigma) + d^\dagger(l,\bz, \sigma)
d(l,\bz, \sigma) ],
\ee
\be
H_{OD1} & = & - [ 2 \kappa a \mu + 4 \kappa^2] 
\sum_l \sum_\sigma \sum_\bz \sum_r {1 \over l}\nonumber \\
&~& ~~~ \Bigg [ 
b^\dagger(l,\bz, \sigma) b(l,\bz+a \br, \sigma) +
b^\dagger(l,\bz, \sigma) b(l,\bz- a \br, \sigma) \nonumber \\
&~&~~~+d^\dagger(l,\bz, \sigma) d(l,\bz+a \br, \sigma) +
d^\dagger(l,\bz, \sigma) d(l,\bz- a \br, \sigma) 
 \Bigg ]
\ee 
and
\be
H_{OD2} & = & - [ { 1 \over 4} - \kappa^2] 
\sum_l \sum_\sigma \sum_\bz \sum_r {1 \over l}\nonumber \\
&~& ~~~ \Bigg [ 
b^\dagger(l,\bz+ a \br, \sigma) b(l,\bz-a \br, \sigma)
+b^\dagger(l,\bz-a \br, \sigma) b(l,\bz+ a \br, \sigma) \nonumber \\
&~&~~~+ d^\dagger(l,\bz+ a \br, \sigma) d(l,\bz-a \br, \sigma)
+d^\dagger(l,\bz-a \br, \sigma) d(l,\bz+ a \br, \sigma) \Bigg ]~.
\ee
 
\subsection{Numerical Investigation}
\subsubsection{Boundary condition}
With the Wilson term added, we do not have decoupled sub-lattices. We have
both nearest neighbor and next-to-nearest neighbor interactions. 
Since with fixed boundary condition, the lowest four eigenvalues are not exactly
degenerate in finite volume, it is difficult to investigate the removal of degeneracy
by the addition of Wilson term.  
With periodic boundary condition, for a
lattice with $2n+1$ lattice points in each transverse direction, 
we identify the $(2n+2)^{th}$
lattice site with the first lattice site. Then for the Hamiltonian
matrix we get the following additional contributions.
\be
H = \begin{pmatrix} {. & . & . & . & ... & . & . & NN & N \cr
              . & . & . & . & ... & . & . & 0 & NN \cr
              . & . & . & . & ....& . & . & . & . \cr
                   ... \cr
                   ... \cr
               . & . & . & . & ....& . & . & . & . \cr
               NN & 0 & . & . & ... & . & . & . & . \cr
               N & NN & . & . & ... & . & . & . & .} 
\end{pmatrix}
\ee
The matrix elements $ NN = -{1 \over 4} + \kappa^2$ 
and $N=-2a\mu\kappa - 4 \kappa^2$.
For a given $n$, the allowed values of $k$ are $k_p= \pm {2 \pi p
\over 2n+1}$, $p=0,1,2, ......$. Thus for $n=3$, we expect multiples
of ${2 \pi \over 7}$ apart from $0$. For $n=5$, apart from $0$,
allowed values of $k$ are multiples of ${2 \pi \over 11}$.  
  
\subsubsection{Numerical results}
Since the Wilson term connects even and odd lattices, the extra
fermions that appear at zero transverse momentum are removed once
Wilson term is added as we now have nearest and next to nearest
neighbor interactions. 
For large $n$, we get the expected spectra but,
numerical results suggest that the finite volume effect is larger for 
small $\kappa$ which is obvious because $\kappa$ is a mass-like parameter. 
For example, with periodic boundary condition, 
for $n=3$, for $ \kappa=1.0, 0.5, 0.4$, we get the expected
harmonics but not for $\kappa=0.1$. The situation is similar for $n=5$. For
$n=10$, expected harmonics emerge even for $\kappa=0.1$ but not for
$\kappa=0.01$. 

\section{Doubling and symmetries on the light front transverse lattice}
Because of  the constraint equation which is inconsistent
with the equal time chiral transformation in the presence of massive
fermions, we should distinguish between chiral symmetry in the equal
time formalism and in the light front formalism.  For example, the free
massive  light front Lagrangian involving only the dynamical degrees
of freedom  is  invariant under $\gamma_5$ transformation. On the
light front, helicity takes over the notion of chirality even in
presence of fermion mass which can be understood in the following way.

   In the two component representation \cite{hz} in the light front formalism,
let  us look at the objects
 $ \psi_L^+$
and $ \psi_R^+$.  We have
\begin{eqnarray}
\psi^+(x) = \pmatrix{ \eta(x) \cr
                    0 \cr}
\end{eqnarray}
with
\begin{eqnarray}
\eta(x) = \pmatrix{ \eta_1(x) \cr
                    \eta_2(x) \cr}
\end{eqnarray}
The projection operators are $ P_R = { 1 \over 2} (1 + \gamma^5)$ 
and $ P_L = { 1 \over 2} (1 -\gamma^5)$ with
\begin{eqnarray}
\gamma^5 = \pmatrix{ \sigma^3 & 0 \cr
                    0 & - \sigma^3 \cr}.
\end{eqnarray}
Then 
\begin{eqnarray}
\psi_R^+= P_R \psi^+ = \pmatrix{ \eta_1 \cr
                     0  \cr
                     0  \cr
                     0}
\end{eqnarray}
and 
\begin{eqnarray}
\psi_L^+= P_L \psi^+ = \pmatrix{ 0 \cr
                     \eta_2  \cr 
                     0  \cr
                     0}. 
\end{eqnarray}
Thus $\psi_R^+=P_R \psi^+$ represents a positive helicity fermion and 
$ \psi_L^+=P_L \psi^+$
represents a negative helicity fermion, even when the fermion is 
{\em massive}.
This makes sense since chirality {\em is} helicity even for a massive
fermion in front form. This is again to be contrasted with the instant
form. In that case the right handed and left handed fields defined by
$ \psi_R= P_R \psi= { 1 \over 2} (1 + \gamma^5) \psi $ and $ 
\psi_L = P_L \psi = { 1 \over 2} (1 -\gamma^5) \psi $ 
contain both positive helicity and negative helicity
states. Only  in the massless limit or in the infinite momentum limit,
 $\psi_R$ becomes the
positive helicity state and $\psi_L$ becomes the negative helicity state. 

As a passing remark, we would like to mention that  in continuum light 
front QCD there is a linear mass term  that allows  for  helicity flip 
interaction.

In  lattice gauge theory in the Euclidean or equal time formalism,
because of reasons connected to anomalies (the standard ABJ anomaly in 
vector-like gauge theories), there  has to be explicit chiral symmetry 
breaking  in the kinetic part of the action or Hamiltonian.
Translated to the light front transverse lattice formalism, this would then require
helicity flip in the kinetic part.   A careful observation of all the
above methods that get rid of fermion doublers on the light front
transverse lattice reveals that this is indeed true.
  
In particular, we draw attention to  the  even-odd helicity flip
transformation
\be
\eta(x_1,x_2) \rightarrow ({\hat\sigma}_1)^{x_1}({\hat\sigma}_2)^{x_2}
\eta(x_1,x_2) \label{eosf}
\ee
that was used in Sec. \ref{lfsf} for spin diagonalization.  
It should also
be clear that the form of the above transformation is not unique in
the sense that  one could exchange ${\hat\sigma}_1$ and 
${\hat\sigma}_2$ and their  exponents $x_1$ and $x_2$ could be 
changed by $\pm 1$.

Note that the Hamiltonians $P^-_{fb}$ given in Eq. (\ref{fb}) and   
$P^-_{w}$ given in Eq. (\ref{w}) that do not exhibit fermion doubling
are not invariant under the
transformation Eq. (\ref{eosf}). On the other hand  the Hamiltonian 
$P^-_{sd}$ given by Eq. (\ref{sd}) that exhibits fermion doubling is invariant under this
 transformation.
 
\section{Summary and Conclusions}
The presence of the constraint equation for fermions on the light front
gives rise to interesting possibilities of formulating fermions on
a transverse lattice. We have studied in detail  the
transverse lattice Hamiltonians resulting from different approaches.
 
In the first approach, forward and backward derivatives are used respectively for $\psi^+$
and $\psi^-$ (or vice versa) so that the resulting Hamiltonian
is Hermitian. There is no fermion doubling. 
The helicity flip (chiral symmetry breaking) term proportional
to the fermion mass in the full light front QCD becomes an irrelevant term  in the
free field limit. With periodic boundary
condition one can get the helicity up and helicity down fermions to be
degenerate for any transverse lattice size $n$. With fixed boundary condition, there is a
splitting between the two states at any $n$ but the splitting vanishes
in the large volume limit.  

In the second approach,  symmetric
derivatives are used for both $\psi^+$ and $\psi^-$. This 
results in four fermion species. This is a consequence of the fact that the
resulting free Hamiltonian has only next to nearest neighbor interactions
and as a result even and odd lattice sites get decoupled. 
One way to remove doublers is to reinterpret them as flavors using
staggered fermion formulation on the light front. In QCD Hamiltonian,
it  generates  irrelevant flavor mixing interactions. However, in
the free field limit, there is no flavor mixing. Another way to remove 
the doublers is to add a Wilson term which generates many extra terms
in the Hamiltonian. In the free field limit, only the  helicity
nonflip terms survive. The  Wilson term 
couples  even and odd sites and  removes the
doublers. Numerically, we found that in small lattice volumes it is preferable to
have not too small values of the Wilson mass $\kappa /a$. 
   
We have tried to understand the fermion doubling in terms of the  symmetries
of the transverse lattice Hamiltonians. We are aware that there are
rigorous theorems and anomaly arguments in the conventional lattice
gauge theories regarding presence of fermion doublers. In standard
lattice gauge theory, some chiral symmetry needs to be broken in the
kinetic part of the action to avoid the doublers. On the light
front, chirality means helicity. For example, a standard Wilson term
which is not invariant under chiral transformations in the
conventional lattice gauge theory, is chirally invariant on the light
front in the free field limit. The question is then why the Wilson term removes the
doublers on the light front transverse lattice. The argument that
there is nonlocality in the longitudinal direction cannot hold
because, in the first place, having nonlocality is not a guarantee for
removing doublers and secondly, there is no nonlocality on the
transverse lattice. One, therefore needs to find a reasoning that
involves the helicity in some way.
We have identified an even-odd helicity flip symmetry of the light front
transverse lattice Hamiltonian, absence of which
means removal of doublers in all the cases we have studied.

Our  interest also lies in studying
finite volume effects on a transverse lattice. As we have emphasized,
 there are important issues to be
understood since (a) any realistic Fock space truncation will force us
to work with relatively small volumes because of the limited availability of 
computing resources and (b) the currently available transverse lattice
formulation uses linear link variables and recovering  continuum limit 
is nontrivial.
We have investigated the effects of fixed and periodic boundary
conditions, which are significant in finite volumes.

Among the many possible extensions of this work, 
it will be  interesting to study the various QCD Hamiltonians 
and to compare the resulting spectra. 
\acknowledgments
One of the authors (A.H) would like to thank James P. Vary for many
helpful discussions. This work is supported in part by the Indo-US
Collaboration project jointly funded by the U.S. National Science
Foundation (INT0137066) and the Department of Science and Technology, India
(DST/INT/US (NSF-RP075)/2001).

\appendix

\section{Forward-backward derivative in conventional lattice theory}
In this appendix we follow Ref.\cite{bd}.
In discretizing the Dirac action in conventional lattice theory the
use of forward or backward derivative for $\partial_{\mu}$ leads to 
non-hermitian action. The hermiticity can be preserved in the following 
way.

In the chiral representation
\be
\gamma^0=\left[ \begin{array}{ll} 0 & -I
\\ -I & 0 \end{array} \right],~~\gamma^i=\left[ \begin{array}{ll} 0 & 
\sigma^i
\\ -\sigma^i & 0 \end{array} \right],~~\gamma^5=\left[ \begin{array}{ll} 
I & 0\\ 0 & -I \end{array} \right].
\ee
The Dirac operator in Minkowski space
\be 
i\gamma^{\mu}\partial_{\mu} \equiv \left[ \begin{array}{ll} 0 & 
-i\sigma^{\mu}\partial_{\mu}
\\ -i{\bar \sigma}^{\mu}\partial_{\mu} & 0 \end{array} \right],
\ee
where, $\sigma^{\mu} =(I,{\bf \sigma}),~~{\bar \sigma}^{\mu} 
=(I,-{\bf \sigma})$. 
For massive Dirac fermions, this leads to the structure
\be
-i\sigma^{\mu}\partial_{\mu}\psi_R -m\psi_L \label{one} \\
-i\sigma^{\mu}\partial_{\mu}\psi_L -m\psi_R. \label{two} 
\ee
For discretization we  replace $\partial_{\mu}$ in Eq. (\ref{one}) by
forward derivative
\be \Delta^f_{\mu}=(\delta_{y,x+\mu}-\delta_{y,x})/a
\ee
and in Eq. (\ref{two}) by backward derivative
\be 
\Delta^b_{\mu}=(\delta_{y,x}-\delta_{y,x-\mu})/a.
\ee
This leads to the structure  
\be
i\gamma^{\mu}\partial_{\mu} - m = i\gamma_{\mu}\Delta^s_{\mu}
-i\gamma_{\mu}\gamma_5\Delta^a_{\mu} - m
\ee
which results in hermitian action.
Here,
\be
\Delta^s_{\mu}&=&(\delta_{y,x+\mu}-\delta_{y,x-\mu})/2a \nonumber\\
\Delta^a_{\mu}&=&(\delta_{y,x+\mu}+\delta_{y,x-\mu}-
2\delta_{y,x})/2a.
\ee
Note that irrelevant helicity nonflip second order derivative term is 
produced in this method of discretization. In contrast, the corresponding 
term in the transverse lattice depends linearly on $m$ and flips helicity.
One can trace this difference to the presence of the constraint equation 
in the light front theory.   


\eject
\vskip .5in
\begin{figure}[hbtp]
\begin{center}
\psfig{figure=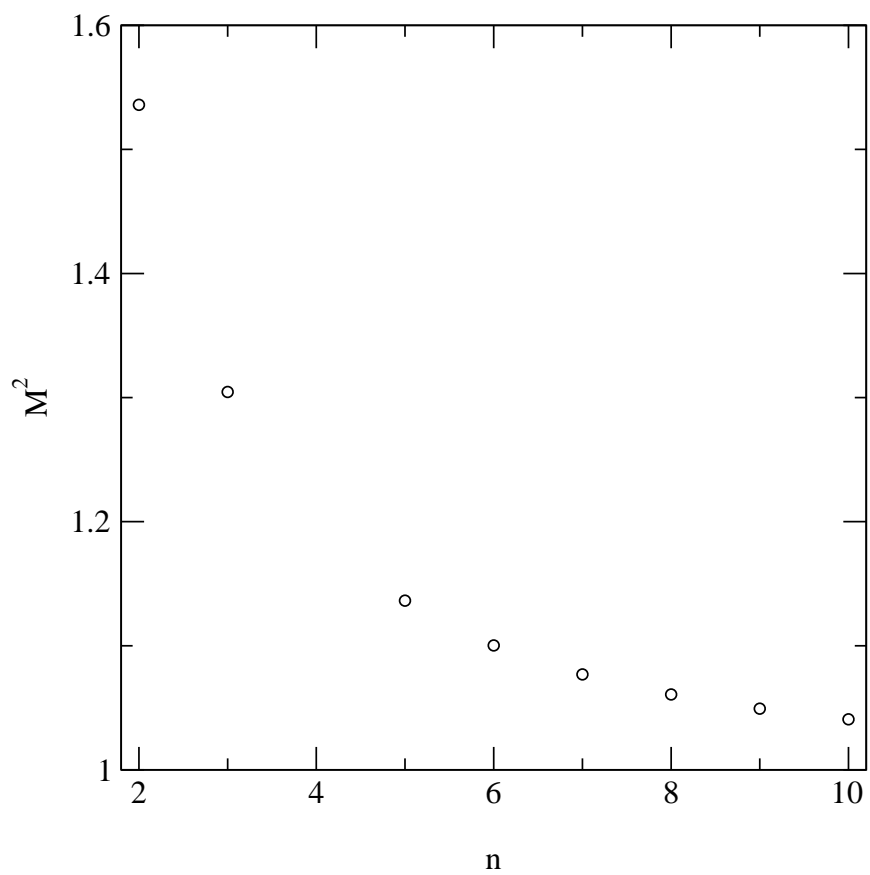,width=6in}
\caption{Ground state eigenvalue versus n.}
\end{center}
\end{figure}
\eject
\begin{figure}[hbtp]
\begin{center}
\psfig{figure=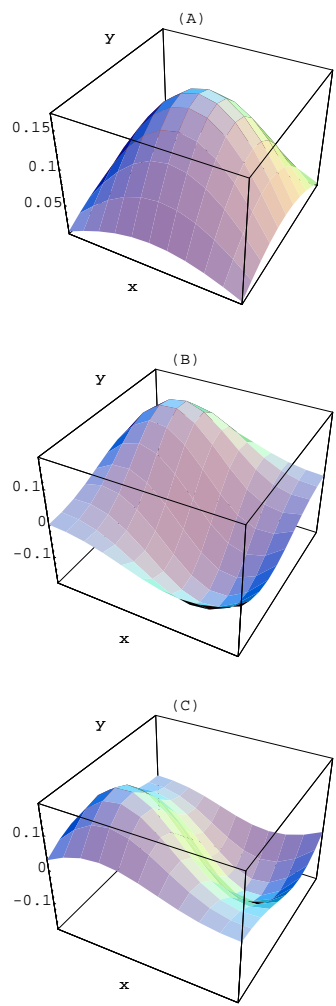,width=4in}
    \caption{Eigenfunctions of first three states for the case of no
doubling. n=5}
\end{center}
\end{figure}

\eject
\begin{figure}[hbtp]
\begin{center}
\psfig{figure=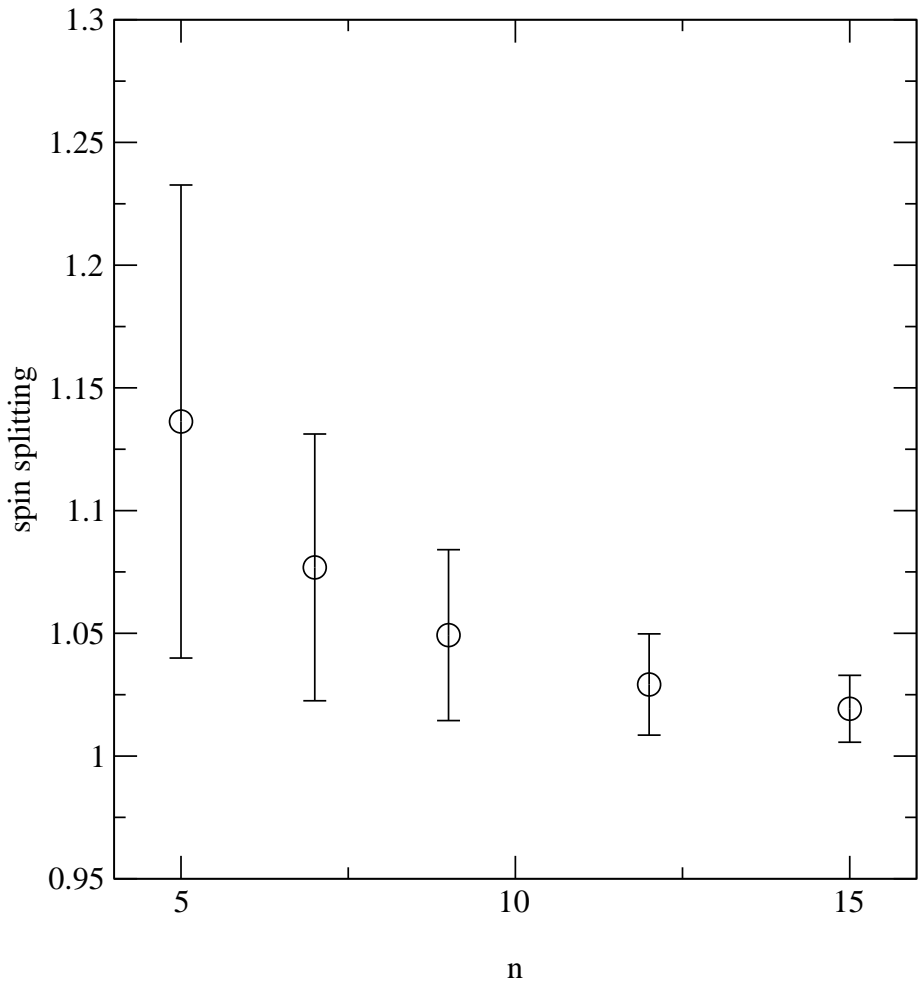,width=6in}
    \caption{Spin splitting of the ground state caused by the spin dependent 
interaction as a function of n.}
\end{center}
\end{figure}
\begin{figure}[hbtp]
\begin{center}
\psfig{figure=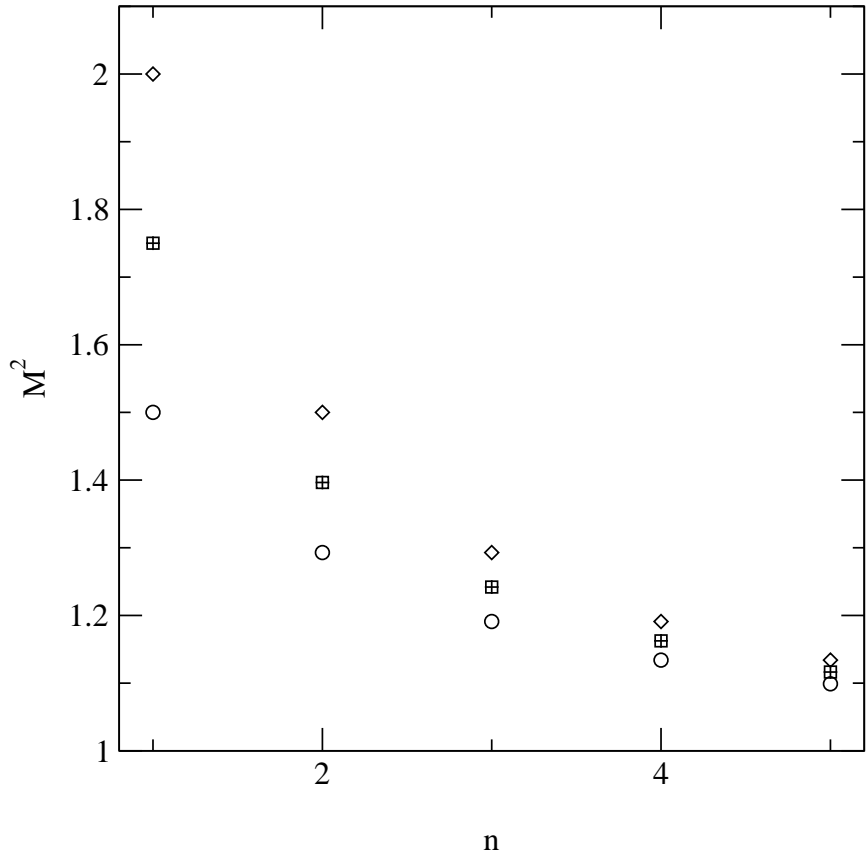,width=6in}

\caption{First four eigenvalues as a function of n.}
\end{center}
\end{figure}
\eject
\begin{figure}[hbtp]
\begin{center}
\psfig{figure=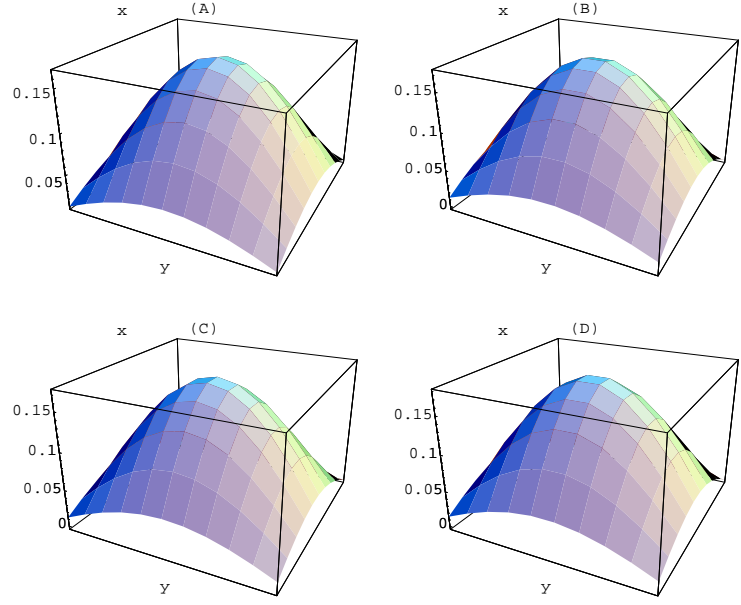,width=4in}
    \caption{Eigenfunctions of first four (degenerate) states for the case
of fermion doubling. n=5}
\end{center}
\end{figure}
\eject
\begin{figure}[hbtp]
\begin{center}
\psfig{figure=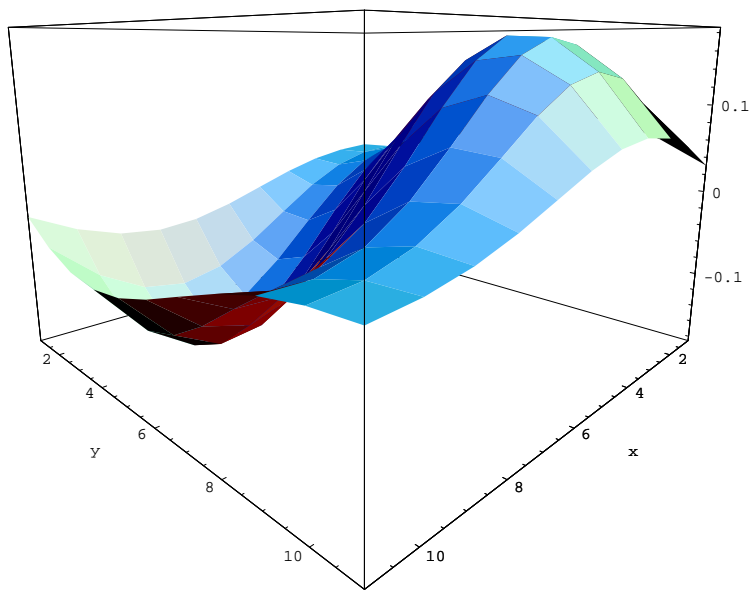,width=4in}
    \caption{Eigenfunction corresponding to the fifth state. n=5}
\end{center}
\end{figure}
\eject

\begin{figure}[hbtp]
\begin{center}
\epsfig{figure=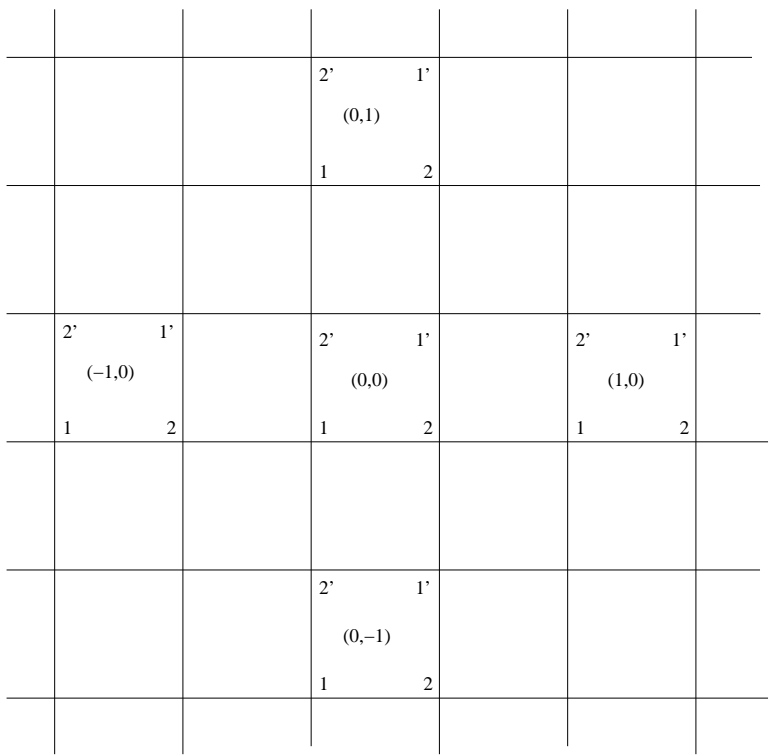,width=4in}
    \caption{Staggered distribution}
\end{center}
\end{figure}
\eject


\end{document}